Graphical abstract:

# Crystallographic evaluation of low cycle fatigue crack growth in a polycrystalline Ni based superalloy

Kaustav Barat[1], Abhijit Ghosh[2], Alok Doharey[3], Shreya Mukherjee[4], Anish Karmakar[5*]

[1] CSIR-National Aerospace Laboratories, Bangalore, Karnataka-560017, India.
[2] Indian Institute of Technology, Indore, Simrol, Madhya Pradesh-453552, India
[3] Indian Institute of Technology (ISM), Dhanbad, Jharkhand 826004, India
[4] Indian Institute of Technology, Kharagpur, West Bengal-721302, India
[5] Indian Institute of Technology, Roorkee, Uttarakhand-247667, India

*Corresponding author email: anish.karmakar@mt.iitr.ac.in*

**Abstract:**
The present work discusses the micro-mechanism of low cycle fatigue (LCF) crack growth in smooth bar specimens of Haynes 282. Two parametric approaches, i.e. crack tip opening angle (CTOA) and maximum tangential stress ($\theta_{MTS}$) have been opted to characterize the cracks. CTOA variations along with a propagating crack, exhibit a non-linear decay followed by a stabilized regime. Mixicity of local $K_I$ and $K_{II}$ fields is directly proportional to $\theta_{MTS}$ and that can be assessed by measuring local deflections. Around the crack, the role of grain incompatibility has been addressed through EBSD and slip transfer analysis. There is a critical bound for Elastic Modulus (EM) and Schmid factor (SF) for the grains favouring subsurface crack propagation, and these values exist beyond a limiting threshold. The SF-EM maps mark the regions of cracked and uncracked grains in the material. The favourable twin-matrix incompatibility of the microstructure has also been identified about the fatigue crack growth and twins in (211) plane is abundant in the cracked region. A detailed slip transfer analysis based on the Luster-Morris parameter (LMP) has been carried out for investigating the interrelation between slip activity, elasto-plastic incompatibility, and grain boundary geometry.

## 1. Introduction:

In recent times, the description of damage draws inspiration from the microstructure of materials. The presence of prior damage-prone microstructural elements and the features leading to heterogeneity has been seen to appear as the probable 'hotspots' for damage initiation. In this work, a detailed mechanism-based study for fatigue cracking is attempted. The material chosen for the study is a polycrystalline Nickel-based superalloy, Haynes 282. For this alloy, the influence of initial microstructure on the high-temperature low cycle fatigue (LCF) has been studied with an assessment of finer (micro/nano level) length scale damage mechanism [1,2]. However, a mesoscale damage description for this alloy has not been studied. In recent times, the mesoscale damage mechanism for LCF cracking is presented in a detailed manner by Stinville et al. [3] focusing on the competing failure modes between the surface to internal crack initiation. However, this study spans from a wide stress range of fatigue, encompassing the LCF (~$10^4$) cycles to HCF (~$10^5$-$10^6$ cycles). Although, limited data and damage descriptions are available in the range of $10^2$-$10^3$ cycles, which generally includes ELCF (Extreme Low cycle fatigue) to early failure regime in LCF. The crystallographic description for a higher cycle range i.e. > $10^6$ cycles is available for Rene′ 88 DT in the literature [4]. Despite temperature intervention, the mesoscale cracking behaviour in HCF is less complex because only elastic incompatibility is playing a key role in the damage progression [5]. Conversely, in LCF damage, the mechanism is more complex because of the interplay of elasticity and plasticity. Also, the role of stress-assisted oxidation enhancing this damage can't be neglected [6].

LCF has been not been treated rigorously and therefore there are some existing gaps in the detailed characterization of secondary cracks and in studying crack initiation and propagation under this regime of fatigue. Where the contribution of plastic strain is much, the cracking can have a crystallographic origin or can generate from arbitrary microstructural inhomogeneities bearing elasto-plastic incompatibilities. Before defining the aim of this work, a detailed state-of-the-art review for existing concepts has been attempted.

## 2. Background:
### 2.1 General aspects of fatigue crack growth:
The process of fatigue crack initiation is random and complex. It greatly depends on the plastic strain amplitude and the environment i.e., temperature, oxidation, partial pressure, and the deformation characteristics of the material (dislocation mobility, dislocation substructure). The generalized process may be subdivided into the following events [7]:

i) Generation of the redundant dislocation density through fatigue hardening or softening to form a cyclically stabilized dislocation population.

ii) Localization of slip through the formation of constrained dislocation substructure, i.e., persistent slip band (PSB).

iii) Interaction of this dislocation substructure with a free surface to produce extrusions and associated intrusions on it.

iv) Stress incompatibility by these intrusions to produce embryonic cracks.

This crack initiation process is hugely modified when an oxide scale is developed on the free surface. Basic mechanisms for oxide-assisted cracking are slip band oxidation and interaction of embryonic cracks with these inclined slip bands [8]. The cracks are often found to be deflected and give rise to a network. The reasons for crack deflections are

(a) If the cracking takes place with 45° orientation along slip lines, it can propagate in mode II and later on when it gets converted to long cracks, propagates horizontally (Figure 1a).

(b) If the cracking initiates in mode I but finds an easier propagation pathway through the oxidized slip bands oriented at 45° (Figure 1b).

(c) If crack initiates and propagates at mode I but later deflects through an oxidized or favorably oriented grain boundaries in an intergranular manner (Figure 1c).

(d) If crack experiences a closure due to deflection in an unfavorably oriented grain ahead of a crack tip (Figure 1d).

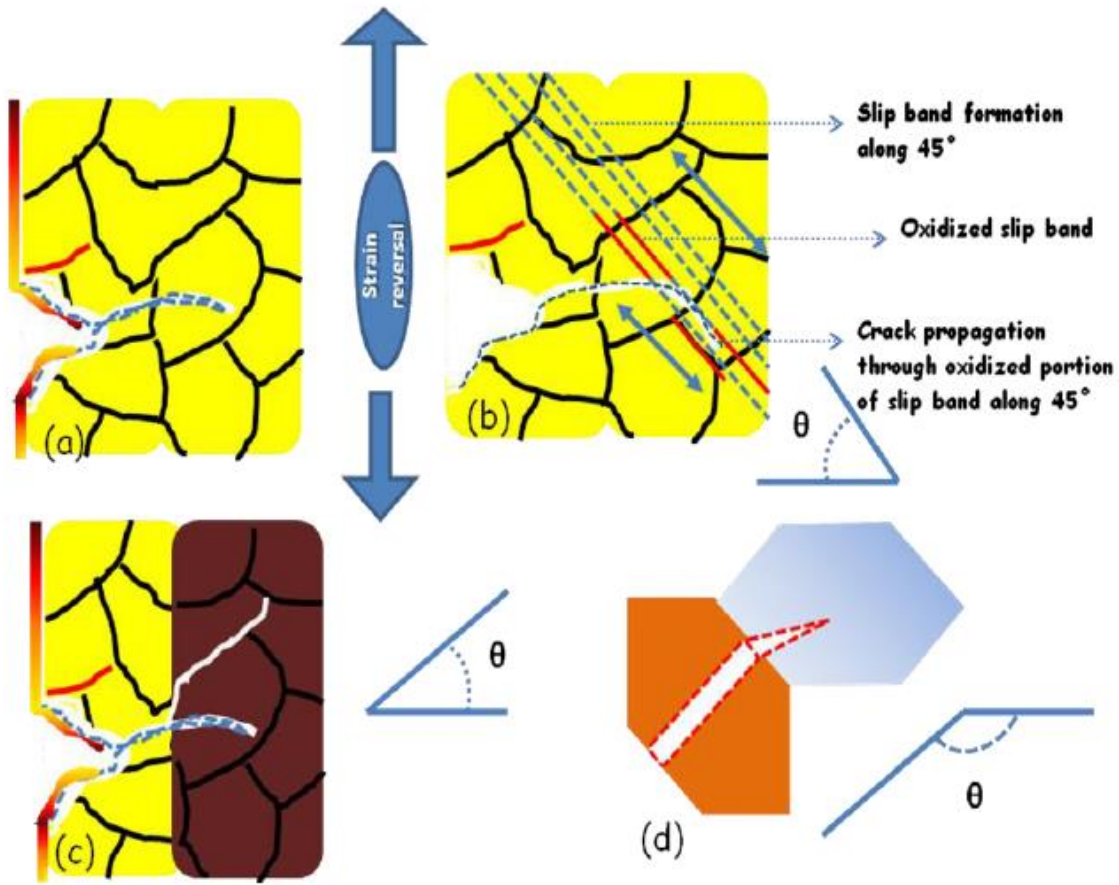

**Figure 1:** Crack deflection caused due to (a) slip band assisted initiation of fatigue cracks (b) mode I initiation and slip band assisted propagation (c) intergranular propagation (d) closure due to unfavourably oriented grain.

The role of different microstructural segments i.e. orientation of individual grains; twin-matrix incompatibility issues as well as ease of slip transfer will be discussed in the following section.

## 2.2 Role of crystallography in crack initiation and propagation:
### 2.2.1 Role of twin boundaries:

As stated in the earlier section, oxidation may have a role in crack initiation but crack propagation and arrest largely depend on internal microstructure where several heterogeneities are associated. Heterogeneities include granular misorientation-based heterogeneity, second phase heterogeneity, grain boundary heterogeneity, and if some existing processing defects are there, heterogeneity also acts as a stress riser. Twin boundaries are reported as crack initiation sites for superalloys [9]. Miao [4] has shown for Rene′ 88 DT, under the HCF regime, the crystallographic facet was formed within a large grain. Large size grains can provide long slip lengths that favor cyclic strain localization. Within the large crack initiation grain, there was a stack of parallel annealing twins. The facet plane was formed in the region close to a twin boundary and intersected the steps of the annealing twin boundary. Gao [10] reported that Σ3 twin boundaries were particularly effective for locally retarding crack propagation because of their large misorientations (60°) in Rene′ 104. So, the results of Miao [4] and Gao [10] contradict each other. Some earlier efforts can also be found in references [11–14] and tabulated in Table 1.

**Table 1**: Crystallographic description of fatigue crack growth in literature

| Author | Material | Temp (°C) | Fatigue type | Fatigue crack initiation site/ plane | Comments |
|---|---|---|---|---|---|
| Vincent and Remy[11] | Mar-M004 | 600 | HCF | (111) (100) | Facet fracture close to crack nucleation threshold |
| Sadananda and Shahinian [12] | Udimet 700 | 25 | HCF | (100) | Faceted fracture close to fatigue threshold |
| Li et al. [13] | Astroloy | 800 | HCF | (111) | Faceted fracture from threshold to striation transition (da/dN>$10^{-5}$ mm/cycle) |
| Miao [4] | Rene′ 88 DT | 593 | HCF | parallel to Σ3 twin boundary (111) | Chevron facet formation at crack initiation and twin mediated nucleation |
| Gao et al. [10] | Rene′ 104 | 25 | HCF | triple junction of random grain boundaries | twin mediated retardation of crack growth |
| Stinville et al. [14] | Rene′ 88 DT | 25 | Stress controlled LCF | favorably oriented grain-twin pair | favorably oriented grain twin pair |

The crystallographic orientations, the angles between the loading direction and the normal of these crystallographic facets were found to vary between 35-65° as determined by Miao et al. [4], where it was seen that all facet normal directions are close to the <111> direction. This result indicates that all these facets were formed on {111} slip planes. Therefore, {111} slip band cracking is the primary mode of subsurface fatigue crack propagation. All these investigations are done on HCF fractured specimens where elastic incompatibility plays a significant role. Things become much more complex when there exists elasto-plastic incompatibility, i.e., in LCF conditions. High elastic anisotropic strain near twin boundary gives rise to higher localized strain field, which also can play a role in crack initiation. So in LCF, the interplay between oxide-assisted crack growth and twin-mediated crack growth becomes competitive, and thus failure occurs.

## 2.2.2 Role of twin-matrix incompatibility:

Twins are the most common microstructural features in Ni-based polycrystalline superalloys. Twins have to play a major role in addressing microstructural integrity issues in these materials [15]. There are a wide range of twin variety and twin boundaries in terms of coherency. Chowdhury et al. [16] have described a wide range of modeling and realization endeavors for fatigue crack growth in highly twinned FCC material. Unique aspects of twins determine their coherency, and that affects the stage I fatigue crack growth locally. For instance, the thickness of the twin changes the fatigue threshold in Ni-Co alloys [17]. The general theorization of stage I fatigue crack growth is that the origin of a fatigue crack is governed by the existing elasto plastic incompatibility strain existing in the material. For the low cycle fatigue process, the strain evolution in tension and compression has been quantified by Stinville et al. [18] and established that localized strains evolve slip bands that are parallel or slightly misoriented with {111} planes parallel to annealing twins.

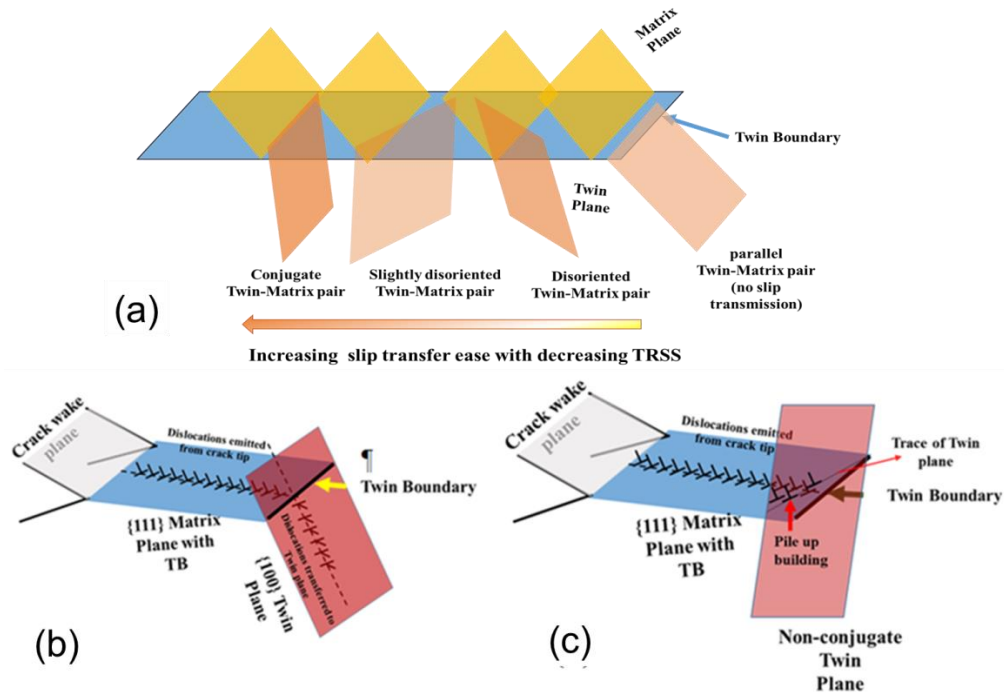

**Figure 2:** (a) Arrangement of Twin-Matrix pairs in deformation process zone. Twin dislocation interaction in (b) coherent twin boundaries (c) twin boundaries having larger twist angle.

In real-life observations, the matrix plane and the twin plane can coexist in four different combinations, as illustrated in Figure 2a. The four combinations are:
(a) Slip plane of matrix parallel to pre-existing twin plane (slip transmission not possible).
(b) Slip plane and twin plane delineated by a boundary of large twist angle (slip transmission is difficult).
(c) Slip plane of matrix having a low misorientation to twin boundary (slip transmission is easier).
(d) Slip plane of matrix sharing a common conjugate twin boundary (slip transmission is easiest).

Therefore, the resolved shear stress across twin boundary (Twin resolved shear stress or TRSS) is highest for configuration (a). It is higher in the case of (b) as compared to (c). It is lowest in the case of (d). Therefore, the susceptibility towards cracking (if twin-matrix interface plays a major role) is highest for the grains having a slip plane parallel to pre-existing twin planes [19].

The crack twin interaction becomes interesting when these different types of interactions take place in a microstructure. Suppose the cracks get initiated from a surface grain and not from a crack twin interface. As we know, the tip of a crack emits dislocations continuously. These dislocations find a suitable slip plane to move along with the advent of crack. Now, different situations can arise when the dislocation emission encounters a twin boundary plane. Different situations are illustrated in Figure 2(b, c).

### 2.2.3 Role of slip transfer in fatigue cracking (Luster-Morris parameter-LMP):

The role of slip has an immense importance on the event of transmission of deformation in microstructural components. Fatigue crack propagation in Ni-base superalloys is one of the processes that can also be viewed from the light of slip transfer. The basic concept of slip transfer is quantified as the geometric incompatibility between two grains. The restriction in slip transfer acts as a stress-raising phenomenon in microstructure and thus results in cracking of the grains. Under pure fatigue loading, the slip bands channelize dislocations from one grain to another; this transmission of dislocation groups can be quantified from the viewpoint of slip transfer. Luster and Morris have shown that the geometric compatibility between two grains depends on the angle between slip plane normal and slip plane directions [20]. The LMP is denoted as maintaining the following relation with the incoming and outgoing slip planes:

$$m' = (\vec{n}_{in}.\vec{n}_{out})(\vec{d}_{in}.\vec{d}_{out}) \qquad \text{eqn. 1}$$

where, $\vec{n}_{in}$ is incoming slip plane normal, $\vec{n}_{out}$ is outgoing slip plane normal, $\vec{d}_{in}$ is incoming slip plane direction and $\vec{d}_{out}$ is outgoing slip plane direction.

In the realm of fatigue deformation, grain boundaries always play an essential role in the propagation and arrest of cracks. The favorable orientation of the grain boundary plane concerning adjacent slip planes also decides the nature of slip transfer. For every fatigue crack, ahead of the crack tip, there lies a reverse plastic zone and monotonic plastic zone. The crack tip emits dislocations creating these two mesoscale zones that are related for the slip transfer across grains and twin boundaries. Twin boundaries are believed to be more resistive against fatigue cracking compared to grain boundaries. Low angle boundaries do facilitate crack propagation by maintaining compliance with the adjacent grains [21]. High angle boundaries and CSL boundaries (low sigma value) are expected to resist fatigue cracking. Σ3 boundaries can inhibit fatigue crack propagation when the trace of the same is parallel to the loading axis. Fatigue crack never propagates along grain boundaries whose stress directions are aligned by angles smaller than 35° from the loading axis [22]. In SUS 316 austenitic stainless steel, the low-energy coherent twin boundaries have the highest fracture strength under monotonic loading. But in fatigue cracking, these boundaries are seen to nucleate fatigue cracks [23–30].

We have made a detailed literature review and discussion on the role of different microstructural components affecting subsurface fatigue crack growth. However, experimental indications for various superalloys are not abundant. Information on high-temperature crack growth with the involvement of other environmental processes is also scarce in the available documents. A detailed study from the fracture mechanics viewpoint and crystallographic viewpoint is necessary to decode the mechanistic and microstructural conditions for crack initiation, propagation, and arrest.

## 3. Experimental details:
### 3.1 Material and Heat treatment:
The material chosen for the present study is a cylindrical forged block of Ni-based superalloy Haynes 282 supplied by GE Energy. It has a nominal chemical composition (wt.%) of: Co 10%, Cr 20%, Mo 8.5%, Fe 1.5%, Si 1.5%, Al 1.5%, Ti 2.1%, C 0.06%, B 0.005% and balance Ni.

The microstructure comprises γ-matrix (FCC) and ordered spherical γ′ ($Ni_3(Al,Ti)$) precipitates. The size range of grain is 130±20 μm.

Two heat treatments in terms of different cooling rates from solutionizing temperature were given to this material; the material was solutionized at 1120°C (above γ′ solvus) for one hour. Two different size distributions of γ′ precipitates were generated; one by slow cooling in a furnace (FC=furnace cooling) giving rise to coarser precipitates and the other by faster cooling in the air (AC=air cooling), giving rise to finer γ′ precipitates.

Blanks of square cross-sections were obtained from the blocks, and after that, LCF specimens were fabricated from those blanks.

### 3.2 LCF testing:
Low Cycle Fatigue (LCF) tests were carried out using INSTRON 8862 servo-electric machine equipped with a high-temperature furnace of maximum temperature 1000 °C. Contact type ceramic extensometer (Gage length 12.5 mm, travel +2.5 mm/-1.25 mm) was used for measuring strain, and K-type thermocouples were wrapped with the specimens for continuous monitoring of the temperature during testing. Three strain amplitudes were chosen (0.4%, 0.6%, and 0.8%) to test two microstructural conditions (furnace cooled and air-cooled) at three test temperatures i.e., 650°C, 700°C, and 760°C. Specimens were tested according to ASTM standard E606 [31] and using a triangular waveform with a constant strain rate of 0.001/sec.

### 3.3 Preparation of sample for characterizing secondary cracks:
The sectioning scheme of the LCF specimen has been shown in Figure 3. The cylindrical gage length portion of the LCF sample has been extracted and subjected to the X-ray tomography system (GE Phoenix, General Electric, Boston, MA) for 3D tomographic imaging of the interior. Visualization and its subsequent 3D rendering of all 2D micro-CT images generated by X-ray scanning have been carried out using VG Studio Max version 2.2 (Volume Graphics, Heidelberg, Germany). All scans were carried out at a 180 kV accelerating voltage, 140 μA current, and an exposure time of 2 s to confirm crack locations.

The LCF tested specimens were cut longitudinally through the midplane. Samples for metallographic examination were prepared, polished and etched in waterless "Glyceregia" reagent (5 ml of $HNO_3$, 15 ml HCl and 10 ml of glycerol). Optical and scanning electron microscopy was carried out in Leica DM 2500M and FEG Quanta 450 (MERLIN), respectively, and micrographs were used to characterize the deflection of cracks and the mechanism for crack formation.

The sectioned midplane from LCF specimens were polished using colloidal silica and thereafter electropolished using 10% perchloric acid in butanol mixture and a bias voltage of -25 V. The EBSD scanning has been done using Cross Beam FIB-FEG Scanning Electron Microscope AURIGA compact manufactured by Carl-Zeiss Microscope GmbH, Germany. A step size of 1 micron and time per step of 0.05 sec has been chosen to record the EBSD data.

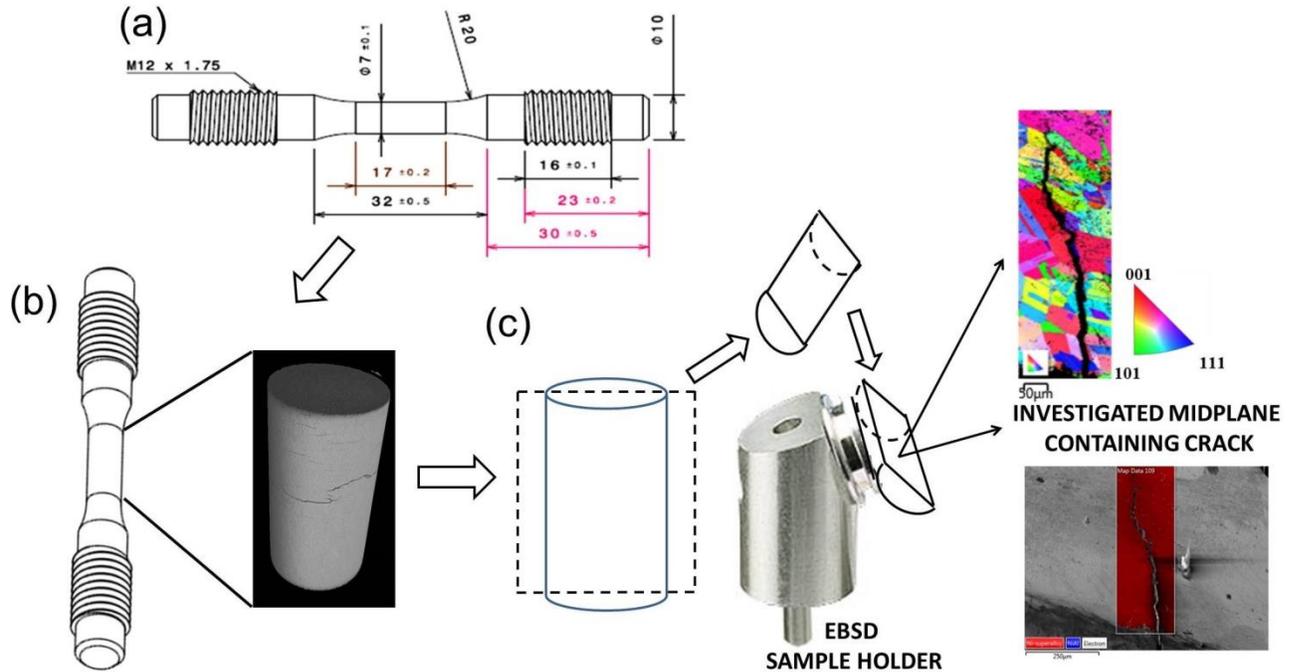

**Figure 3:** Sequence of sample preparation (a) Drawing of LCF sample with isometric view of the gauge length from where sample has to be extracted, (b) Tomographic side view of the sample, (c) scheme for the midplane sectioning, polished midplane containing crack and EBSD characterization of the microstructure around the crack.

## 4. Result and discussion:
### 4.1 LCF test results: Strain amplitude evolution:

Cyclic hardening/ softening behavior indicated by variation in stress amplitude for both the microstructures (furnace cooled and air-cooled) has been analyzed. Plots for different test temperatures and different microstructures are shown in Figure 4. It is observed that while at 650°C, predominantly hardening occurs, but at 760°C, mainly softening occurs. Another interesting point to note is that at 700°C, among air-cooled (AC) samples, while samples tested at lower strain amplitude (0.4%) show softening, samples tested at higher SA (0.8%) show hardening during HTLCF tests. This suggests that dislocation-microstructure interaction varies with test temperature and strain amplitude and an effect of precipitate size is thus obvious on the material's response to cyclic loading. Cyclic hardening phenomenon can be attributed mainly to the resistances to anti-phase boundary (APB) shearing, cross slip, Orowan looping around γ′, activation of multiple slip systems
888
88888
88
8
8
8
8

and interaction of the forest dislocations with slip bands. Cyclic softening can be ascribed to the shearing of γ′ precipitates and highly localized shearing of APBs on glide plane.

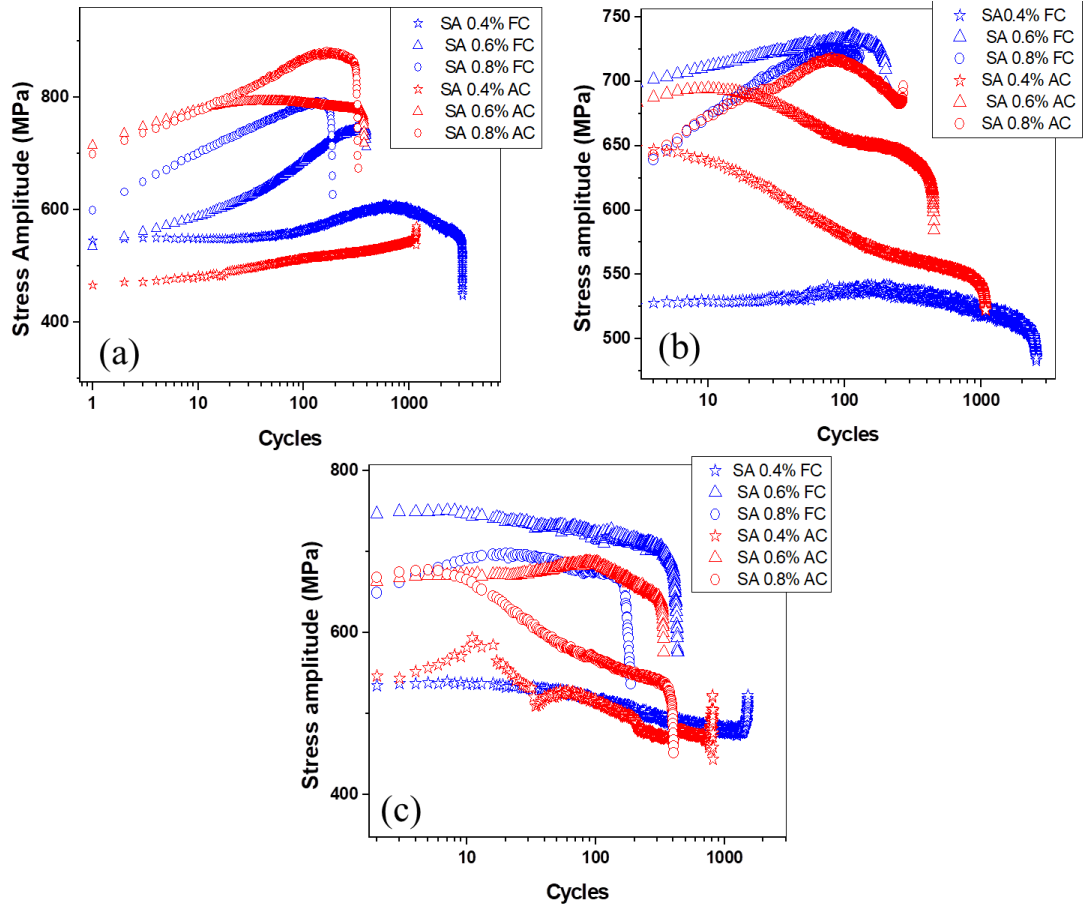

**Figure 4:** Stress amplitude vs. no of cycles plots for (a) 650°C (b) 700°C (c) 760°C; FC-Furnace cooled, AC-Air cooled.

## 4.2 CTOA analyses and defining a stability ratio criterion for crack paths:

Analysis methods and procedures for this has been elaborated in Appendix-1. Figure 5 (a) and (b) show optical micrographs containing secondary cracks. Corresponding CTOA vs. a graphs are shown in Figure 5 (c) and (d) respectively. In each of these graphs non-constant and constant regions have been identified. Such graphs and related average values of non-constant/constant lengths of crack propagations have been measured. These data for all the LCF test conditions (different strain amplitudes, test temperatures and microstructures) are shown in Figure 5e. The results show the stability statistics of cracking. Following specific observations are made:

- Most of the cracks have a non-constant to constant crack length ratio lower than 1.
- With increasing strain amplitude non-constant/constant crack length ratio increases at intermediate temperature (700°C).
- Statistically, also at higher strain amplitudes, large fractions of cracks have been seen to have grown in an unstable manner.

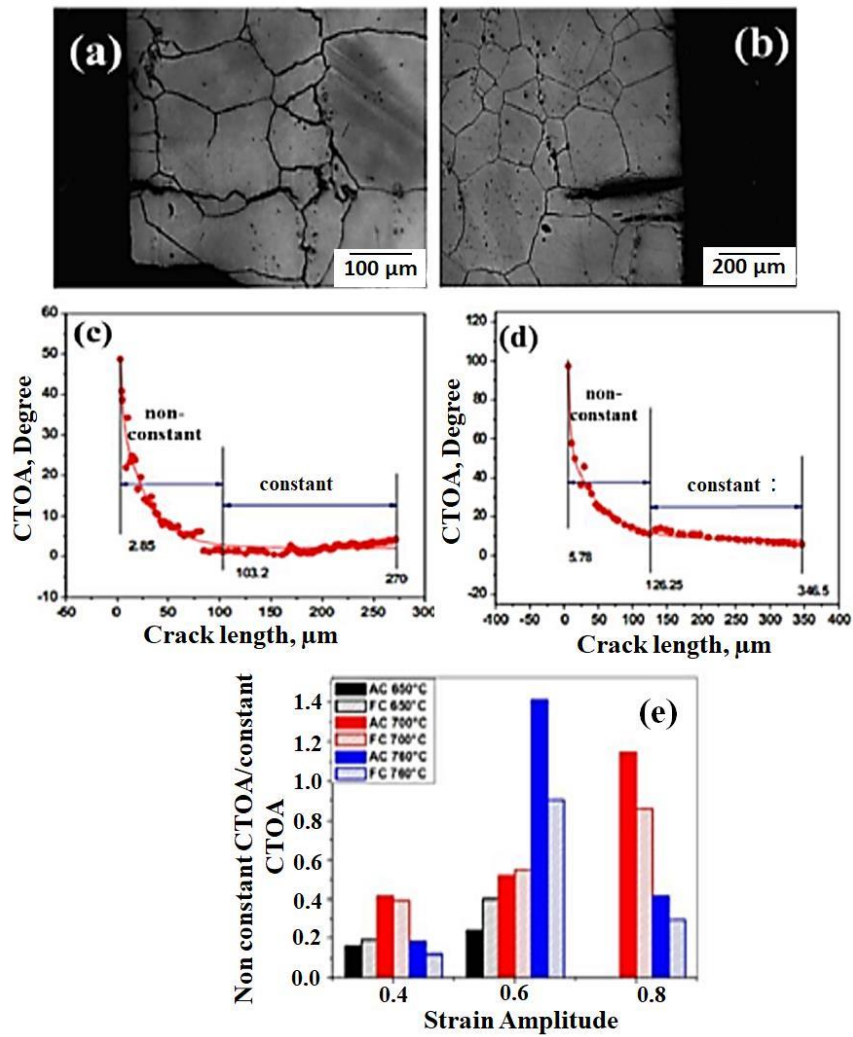

**Figure 5:** (a,b) Optical micrographs of two sub-critical cracks captured from the LCF tested specimens, (c,d) determination of non-constant and constant CTOA portions associated with crack growth (e) variation of average non constant CTOA/constant CTOA ratio at different strain amplitudes.

- For samples tested at higher temperatures (700°C and 760°C) air cooled (AC) samples have shown a higher unstable/stable crack length ratio compared to the furnace cooled (FC) samples.
- There are inflection points in the stable region of the CTOA vs. **a** plots as marked in the Figure 5 (c) and (d). These bear the signature of re-initiation of cracking process.

This non-constant/constant ratio has been formulated for the parametric characterization of secondary cracks. These cracks generally have random initiation sites on the specimen surface and no probe can be attached to measure in-situ growth and meandering characteristics of these cracks. Non-constant/constant ratio gives a platform to comment on a blind process of subsurface crack growth. Higher constant length of stable cracks at high/low strain amplitudes (0.4% and 0.8%) gives rise to the notion that due to higher stress intensity factor (for high strain amplitude cases) and longer time for crack propagation (in low strain amplitude cases) we have higher constant CTOA regime.

Non-constant CTOA regime supersedes stable CTOA regime when there is an intermediate stress intensity factor accompanied by intermediate incubation time for crack propagation and arrest.

### 4.3 Crack deflection and root causes:

Instead of being subjected to a purely Mode I loading, several instances of crack deflections have been seen in the present investigation. According to Suresh et al. [32] the deflecting nature can be categorized in different types, i.e. (i) Kinked cracks, (ii) forked or bifurcated cracks, (iii) doubly kinked cracks, (iv) zigzag cracks, (v) twisted cracks. Figure 6 presents a schematic describing all the crack deflection phenomena like (a) crack meandering, (b) doubly kinked crack, (c) kinked crack with branching (d) zigzag crack (e) 45° deflection with cracking plane and (f) crack tip bifurcation.

From the optical micrographs presented here, we have evidenced all four types of cracks except twist cracking in the present investigation. The basic reasons for crack deflections are:

(a) If the cracking takes place along 45° orientation along slip lines, it can propagate in mode II and later on when it gets converted to long cracks, propagates horizontally (Figure 6a).

(b) If the cracking initiates in mode I but finds an easier propagation pathway through the oxidized slip bands oriented at 45° (Figure 6b).

(c) If crack initiates and propagates at mode I but later deflects through oxidized or favorably oriented grain boundaries in an intergranular manner (Figure 6c).

(d) If crack experiences a closure because of deflection in an unfavorably oriented grain ahead of a crack tip (Figure 6d).

These are the root causes of crack deflection and there always exist crystallographic reasons for these deflections. Detailed discussions about some aspects are incorporated in section 4.6.

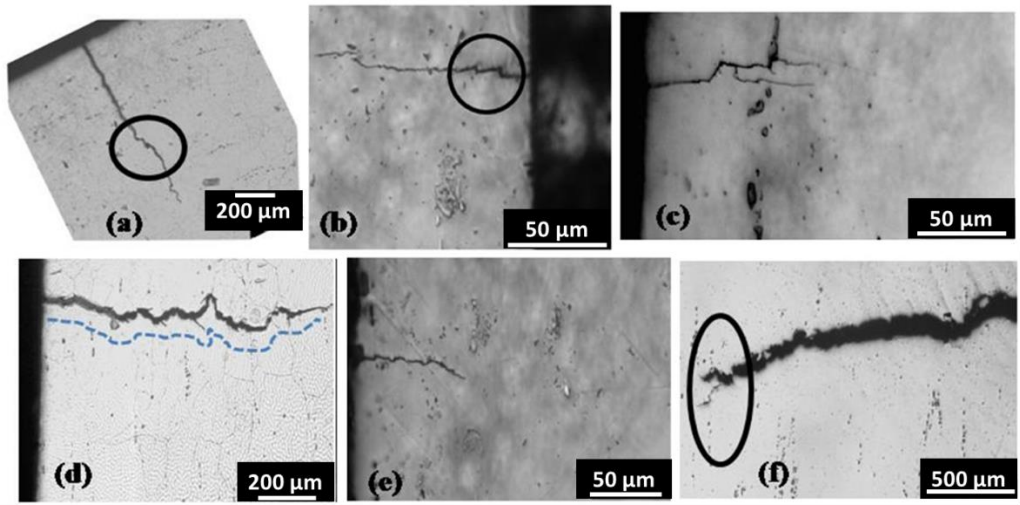

**Figure 6:** (a) Meandered crack (b) Doubly kinked crack (c) Kinked crack with branching (d) zigzag crack (e) 45 ⁰ deflection (f) crack tip bifurcation.

### 4.4 Mode mixicity in subsurface cracking under LCF:

According to the maximum tangential stress criterion for crack propagation, experimentally measured deflection angle ($\theta_{deflection}$) is equal to the angle for maximum tangential stress ($\theta_{MTS}$) [33]. $\theta_{MTS}$ is related to the local stress intensity factors in different modes i.e. $K_I$ and $K_{II}$ (Mode I corresponds to crack opening mode and mode II corresponds to the in-plane shear mode) by the following equation:

$$\theta_{deflection} = \theta_{MTS} = \tan^{-1}\left(\frac{K_{II}}{K_I}\right) \qquad \text{eqn. 2}$$

Hence, the measured deflection angle can be considered as a parameter representing mode mixicity of different cracks. This mode mixicity parameter actually gets dictated by geometric planes corresponding to minimum stress intensity factor, minimum shear modulus, and crystallographic constraints [34].

Figure 7(a) and (b) show the deflection angles of mixed mode cracks in samples tested at 700°C at 0.6% and 0.8% strain amplitudes respectively. Figure 7 (c) and (d) present the values of the mode mixicity parameter ($\theta_{MTS}$) for different conditions of LCF tested samples.

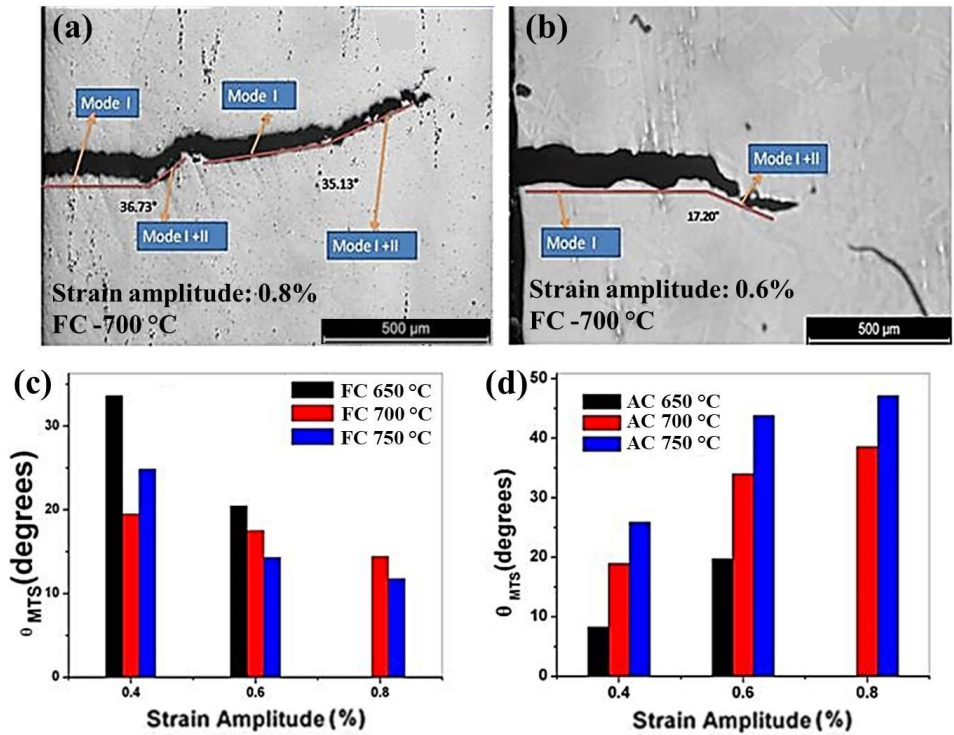

**Figure 7:** Mixed mode cracking in 700°C at (a) 0.6%, (b) 0.8%, (c) statistical distribution of experimentally measured θ $_{MTS}$ with strain amplitude (in %) for (c) Furnace cooled microstructure and (d) for air cooled microstructure.

Especially, for furnace cooled microstructure for all strain amplitudes (0.4%, 0.6% and 0.8%) mode mixicity has not been seen to exceed 25° and air cooled condition, mode mixicity has not exceeded 40° as shown in Figure 7(c, d). At 0.6% and 0.8% strain amplitudes, we have observed almost comparable mode mixicity for air cooled specimens for 700°C and 760°C tested samples. According to our observation, mode mixicity is typically a microstructurally dependent parameter. This is consistent with the observations of Nalla et al. [33] and Campbell and Ritchie [34] who based on their various works on titanium alloys reported local microstructure affects the mixed-mode cracking to a great extent.

## 4.5 Physical mechanism of crack formation and propagation:
### 4.5.1 Stages of crack formation:

Oxidation also plays a prominent role in crack initiation and propagation. A detailed observation has been made on the basis of oxide scale formation, stress assisted grain boundary oxidation (SAGBO) and conversion of short or small cracks to meso and macrocracks. The detailed analysis is based on the arrested secondary cracks present in the failed specimens. Different stages of secondary crack formation have been detected and they are sequenced chronologically as below (Figure 8 (a-h):

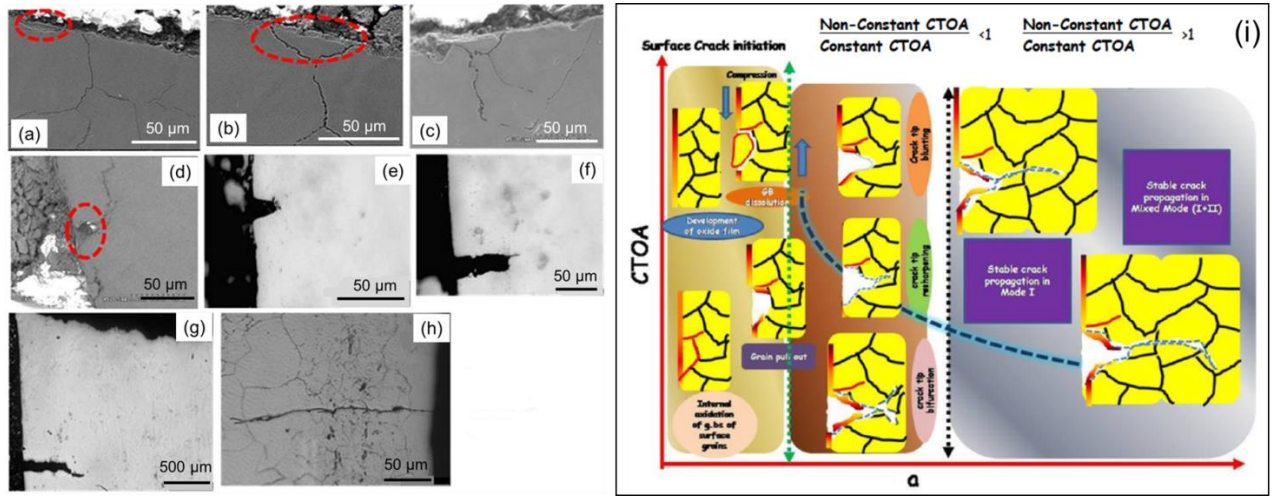

**Figure 8:** (a)-(h) Sequence and microstructural modes of crack propagation in Haynes 282 as envisaged in the present study. (i) A meso-mechanism map of oxide scale assisted crack propagation under High temperature LCF conditions

- **Stage 1: Oxide scale formation and spallation:**
  The occurrence of oxide scale building, spallation and rebuilding takes place in the first few cycles of low cycle fatigue. Oxide scale actually formed in the initial cycles and thereafter goes through a dynamic process of spallation and reforming. Cracks formed in the oxide scales are found to get arrested at the scale specimen interface (Figure 8a).

- **Stage 2: Internal oxidation of grain boundaries and formation of microstructural notch because of grain pull out:**
  Internal oxidation of grain boundaries takes place when there is a preferential attack. The preferential oxidation of grain boundaries facilitates decohesion and pushing off the surface grain from the sample. It gives rise to a microstructural notch that starts the fatigue crack (Figure 8 (b-d)).

- **Stage 3: Notch tip resharpening and bifurcation:**
  After the formation of microstructural notch, it acts as a crack starter. The notches formed due to grain pullout have been found to represent a blunt notch (Figure 8e). With cycling, it gets resharpened giving rise to a propagating crack (Figure 8f). This propagating crack concomitantly interacts with grain boundaries and slip bands. While propagating through grain boundary it gets branched. When the propagating crack interacts with slip bands aligned at 45° it acquires a forked or bifurcated shape.

- **Stage 4: Crack propagation and deflection:**
  The fatigue cracks investigated in this study have been found to spend their life as short cracks in stable regime and then transform to long cracks. We can see some fluctuations in CTOA vs. crack length curve and that has been seen due to re-initiation of a crack. The crack has been seen to propagate under mixed mode after initial propagation in mode I (Figure 8g). This mode transition has been seen in many cases and another reportable observation is that this transition in all cases ends in crack closure (Figure 8h).

### 4.5.2: Meso-scale mechanism map for secondary cracking:

Figure 8(i) represents a mechanism map, describing the whole process of initiation and propagation based on our observations of various type of cracking. In the same figure, CTOA vs. **a** curve is superimposed. After initiation, in the subsequent loading cycle, the microstructural blunt notch undergoes further blunting with the application of tensile straining. Again, after compression, the blunt crack tip gets sharpened and the crack propagation starts. For the tendency of crack to propagate at the 45° angle, the crack tip gets bifurcated and after that, the unstable crack propagation starts. However, propagation gets interrupted by oxide induced closure.

### 4.6 Crystallographic analysis of microstructure around secondary cracks:

A detailed study of secondary non-propagating cracks, detrimental cracks responsible for ultimate failure and the surrounding microstructure has been analyzed. This section aims to unravel the crystallographic participation of the crack initiation, propagation and arrest process.

### 4.6.1 Crack initiation: The crystallography of the local microstructure

In the present study, we have selected a specimen having a facet at the edge of the specimen and the detrimental crack is supposed to get initiated from a surface grain. The testing detail for this sample as per nomenclature is S4, (HTLCFR 6T700 FC, tested at strain amplitude 0.6% at 700 °C, microstructure is Furnace cooled). In Figure 9, the EBSD results have been presented.

Figure 9a is depicting the EBSD indexed zone perpendicular to the crack propagation plane. The micrographs from the EBSD analysis depicted a population of large and small grains having annealing twins with 89 % of high angle grain boundaries (HAGBs), Figure 9b. The twin boundaries are clearly visible within the grains. Kernel average misorientation (KAM) map of the scanned portions clearly showing more strain in the grains situated near the main fracture plane, as showed by the colour legends in Figure 9c, where preferably more strain is showed by the yellow/green colour compared to the blue regions.

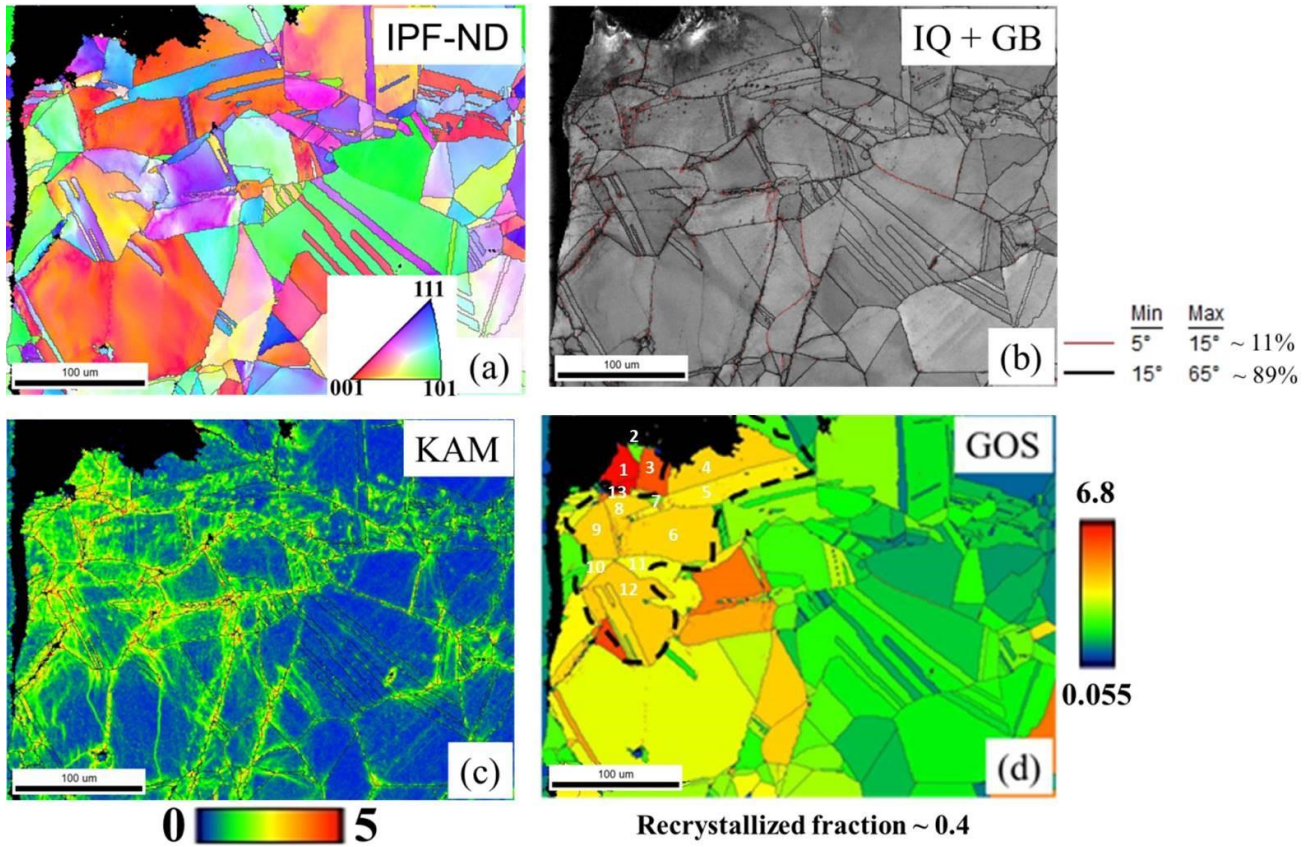

**Figure 9:** EBSD analysis of the crack initiator grains at the facet plane. (a) Inverse pole figure (ND) with color legend in the inset, (b) Image quality map superimposed with grain boundaries (IQ+GB), (c) kernel average misorientation (KAM) and (d) grain orientation spread (GOS) map of the same.

The strain is selectively partitioned in the grain boundaries of the coarse grains (average grain size > 90 µm), Figure 9c. It is quite clear that more strain is partitioned towards the domains nearer to the main fracture plane irrespective of the grain size; Figure 9c. Grain orientation spread (GOS) map of the selected regions exhibited the recrystallized fractions of the grain regions. Figure 9d. The GOS value less than 1° is considered to be the recrystallized fractions of any polycrystalline material which is around 40 % for the current investigated region, Figure 9d.

Multiple grains in superalloys have been seen to construct the facet. Often these grain clusters are referred as supergrains [4]. Here, the facet contains a grain cluster having a similar grain orientation spread (GOS value) i.e. encircled in black dotted line in Figure 9d. Each grain of this supergrain cluster contains at least one twin boundary.

At first slip, planes are transformed from the crystal reference frame to sample reference frame for a given orientation of crystal with the help of following equation.

$$C_s = g_1^{-1} \cdot C_c \qquad \text{eqn. 3}$$

*Cc* represents the plane-normal in crystal reference frame, *Cs* represents the same in sample reference frame, $g_1$ represents the orientation of any matrix grain in the sample reference frame.

In a FCC system, there are 4 possible primary slip planes. Therefore, there can be 16 angles between them. But the combination of slip systems of the neighbouring grains provides the minimum angle that should be selected for the slip transfer. The same method has been adopted in calculation of the effective grain size in ferritic steel [29,35–37].

For example, the boundary between grain1 and grain2 has a high misorientation angle, 58.39°. But, the lowest possible angle between the (111) slip planes of those grains will be around 5° which is between the {11-1} of grain1 and {-1-11} of grain 2. Since this angle is less than 15° deg, therefore slip transfer can take place.

**Table 2:** Calculated Minimum angle for slip transfer, slip resistance in neighborhood grains are stated in bold letters, grains not in neighborhood are stated in red color.

| Grain Id | 1 | 2 | 3 | 4 | 5 | 6 | 7 | 8 | 9 | 10 | 11 | 12 | 13 |
|---|---|---|---|---|---|---|---|---|---|---|---|---|---|
| 1 | 0 | 5.0532 | **18.159** | 2.1842 | 4.0381 | 3.7153 | 22.498 | 8.9952 | 20.017 | 26.502 | 8.1678 | 20.759 | 8.5731 |
| 2 | 5.0532 | 0 | 10.767 | 10.331 | 10.936 | 9.7601 | 12.790 | 12.938 | 22.842 | 25.892 | 14.756 | 19.231 | 20.658 |
| 3 | **18.159** | 10.767 | 0 | 4.2887 | 21.149 | 7.4674 | 2.7358 | 7.9645 | 13.294 | 14.722 | 8.4952 | 8.1878 | 18.884 |
| 4 | 2.1842 | 10.331 | 4.2887 | 0 | 2.5965 | 1.7744 | **20.426** | 3.6549 | 14.545 | 25.531 | 8.1405 | 10.171 | 9.0696 |
| 5 | 4.0381 | 10.936 | 21.149 | 2.5965 | 0 | 3.5167 | **17.912** | 8.4347 | 16.329 | 18.120 | 10.639 | 18.471 | 11.658 |
| 6 | 3.7153 | 9.7601 | 7.4674 | 1.7744 | 3.5167 | 0 | **18.988** | 5.2520 | **17.933** | 24.646 | 7.3062 | 8.1827 | 8.6027 |
| 7 | 22.498 | 12.790 | 2.7358 | **20.426** | 17.912 | 18.988 | 0 | 14.606 | 20.387 | 13.460 | 23.998 | 21.810 | 15.949 |
| 8 | 8.9952 | 12.938 | 7.9645 | 3.6549 | 8.4347 | 5.2520 | 14.606 | 0 | **17.591** | 26.325 | 7.0330 | 13.268 | 9.1299 |
| 9 | 20.017 | 22.842 | 13.294 | 14.545 | 16.329 | **17.933** | 20.387 | **17.591** | 0 | 12.186 | **23.701** | **19.187** | 2.9912 |
| 10 | 26.502 | 25.892 | 14.722 | 25.531 | 18.120 | **24.646** | 13.460 | **26.325** | 12.186 | 0 | **16.728** | 8.8261 | 22.120 |
| 11 | 8.1678 | 14.756 | 8.4952 | 8.1405 | 10.639 | 7.3062 | 23.998 | 7.0330 | **23.701** | **16.728** | 0 | 11.153 | 2.1467 |
| 12 | 20.759 | 19.231 | 8.1878 | 10.171 | 18.471 | 8.1827 | 21.810 | 13.268 | **19.187** | 8.8261 | 11.153 | 0 | 21.742 |
| 13 | 8.573 | 20.658 | 18.884 | 9.0696 | 11.658 | 8.6027 | 15.949 | 9.1299 | 2.9912 | 22.120 | 2.1467 | 21.742 | 0 |

Similarly, the feasibility of the slip transfer for all other boundaries has been evaluated by the above procedure. The threshold angle for slip transfer has been assumed to be 15° in all cases. Table 2 is illustrating the angles considering all the grains (portion surrounded by a black dotted line in Figure 9d and their mutual minimal angles within slip planes.

The above table explicitly considers all the possibilities of slip transfer with all grain combinations, as shown in Figure 9d. But practically, the grains those are not in geometric neighborhood cannot be considered and highlighted in red colour. If the mutual angle for slip transfer between neighboring grains is more than 15°, those grains are expected to resist slip transfer and stated in black bold letters in Table 2.

## 4.6.2 Description of crack propagation: Crystallographic participation

Two different samples have been chosen for crack propagation study. They are Sample 1 (S4) and Sample 6 (S6). The testing details for S4 and S6 are

S4: HTLCFR 6T700 FC – 1, tested at Strain amplitude 0.6% at 700°C, microstructure is Furnace cooled

S6: HTLCFR8T700 FC – 2, tested at Strain amplitude 0.8% at 700°C, microstructure is Furnace cooled

Detailed analysis of the EBSD data of S4 is presented in Figure 10. As envisaged from Figure 10a, almost more than 95% of boundaries are high angle grain boundaries with a misorientation of > 15°, as indicated by black lines in Figure 10a Inverse pole figure map manifesting the random orientations of the samples. Within the randomness, near about 40% grains are having an orientation of <111> || ND, with a tolerance angle of 20°, Figure 10b. There is not much difference of strain in the whole matrix considering both the zones nearer and further from the crack path, Figure 10c. All possible CSL boundaries in FCC system have been highlighted in Figure 10d. The area percent of Σ3 type of CSL boundaries are dominating (~ 30%) over the others. Around the crack portion, there are twins which are having <211> orientations (~ 16%).

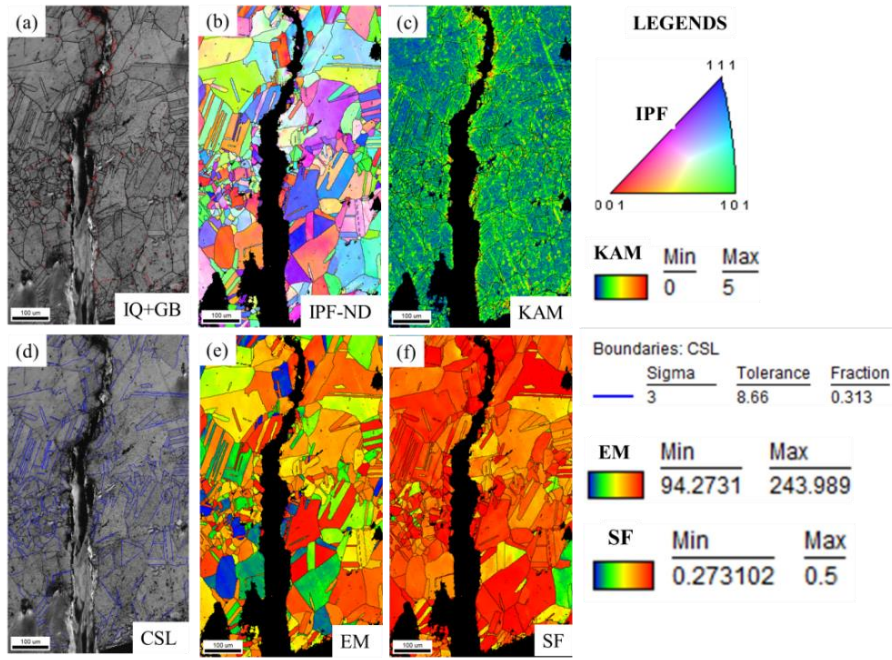

**Figure 10:** (a) Image quality with grain boundary misorientation (IQ+GB), (b) Inverse pole figure regarding sample normal (IPF-ND), (c) Kernel Average Misorientation (KAM), (d) CSL boundary, (e) Elastic Modulus (EM) and (f) Schmid factor (SF) map of S4 sample. The legends are showing the accurate representation of the colors in the respective maps.

Figure 10(e, f) show that the elastic and plastic incompatibility maps. The elastic modulus map shows the elastic modulus of different grains along crack path and Schmid factor map shows the distribution of Schmid factor along the crack path. The elastic modulus values have been taken from Neighbours et.al [38] as $C_{11}$ =253 GPa, $C_{12}$=152 GPa and $C_{44}$=124 GPa.

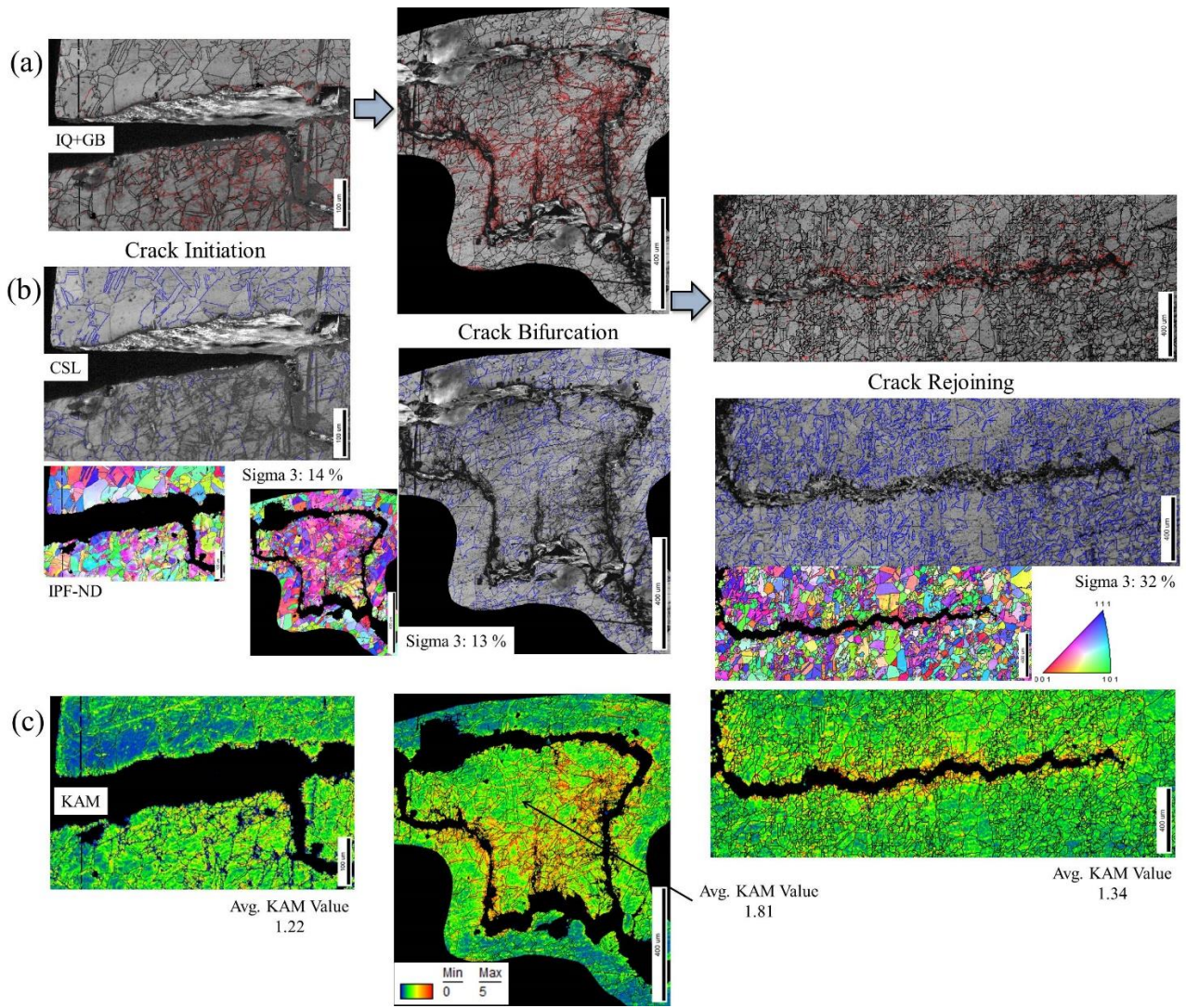

**Figure 11:** Crack initiation, bifurcation and rejoining phenomena of S6 sample. Image quality maps (a) are showing the high angle (θ>15°-black lines) and low angle (5°<θ<15°-red line) grain boundaries as well as Σ3 CSL boundaries (blue lines) (b). IPF-ND of the same samples are depicting random texture in the matrix (b), KAM maps are represented in (c).

In S6, some unforeseen characteristics have been observed. The crack initiated from surface, got bifurcated and created an intermediate island and then get rejoined. After rejoining it propagated through a very fine, almost recrystallized grain zone and arrested subsequently. Therefore, S6 is having three regions to be analyzed i.e. (i) First branch, from initiation to bifurcation denoted as S6-1 (ii) second branch, from bifurcation to rejoining (S6-2) and (iii) Third branch, from rejoining to arrest (S6-3). Critical observations from this sample are as below:

Figure 11a represents a clear abundance of low angle boundaries (red lines in IQ+GB map) in the crack wake zone. As the annealing twins were prevalent in the parent microstructure, an increase in Σ3 boundaries were prominent in S6 (Figure 11b). Unlike the low angle boundaries, the Σ3 boundaries are randomly distributed in the microstructure. The grain boundary distribution gives the impression of the crack tip plastic zone promotes the low angle boundaries, but a competitive process of crack tip stress relieving and dislocation build-up was dynamically embedded within the entire process. The heterogeneity of strain localization in crack tip and wake region can also be observed and calculated from the KAM value (Figure 11c). The island region within two branches of cracks (S6-2) is heavily strained. Twin originator grain pair is involved in creating local strain incompatibility and initiated the fatigue crack. Strain has been preferentially partitioned to non-twinned grains. This strain gives rise to dynamic recrystallization in non-twinned grain and brings down the partitioned strain. The relation between twinning and recrystallization for superalloys has been described in [39] in a detailed manner. Recrystallized grain cluster with random orientations and grain clusters with large twin boundary fraction probably arrested the crack (S6-3).

### 4.6.3 Elasto plastic incompatibility:

Partition of elasto plastic strain is the main reason to define the damage accumulation in materials. In superalloys, the commonly known stress raisers are inclusions [40], twin grain boundaries, slip line-annealing twin intersections [13] etc. Detailed high resolution DIC studies have been done by Stinville et al. [41] in order to map the strain fields and it was observed that twin boundaries are the zones to contribute to the maximum incompatibility strains in the materials.

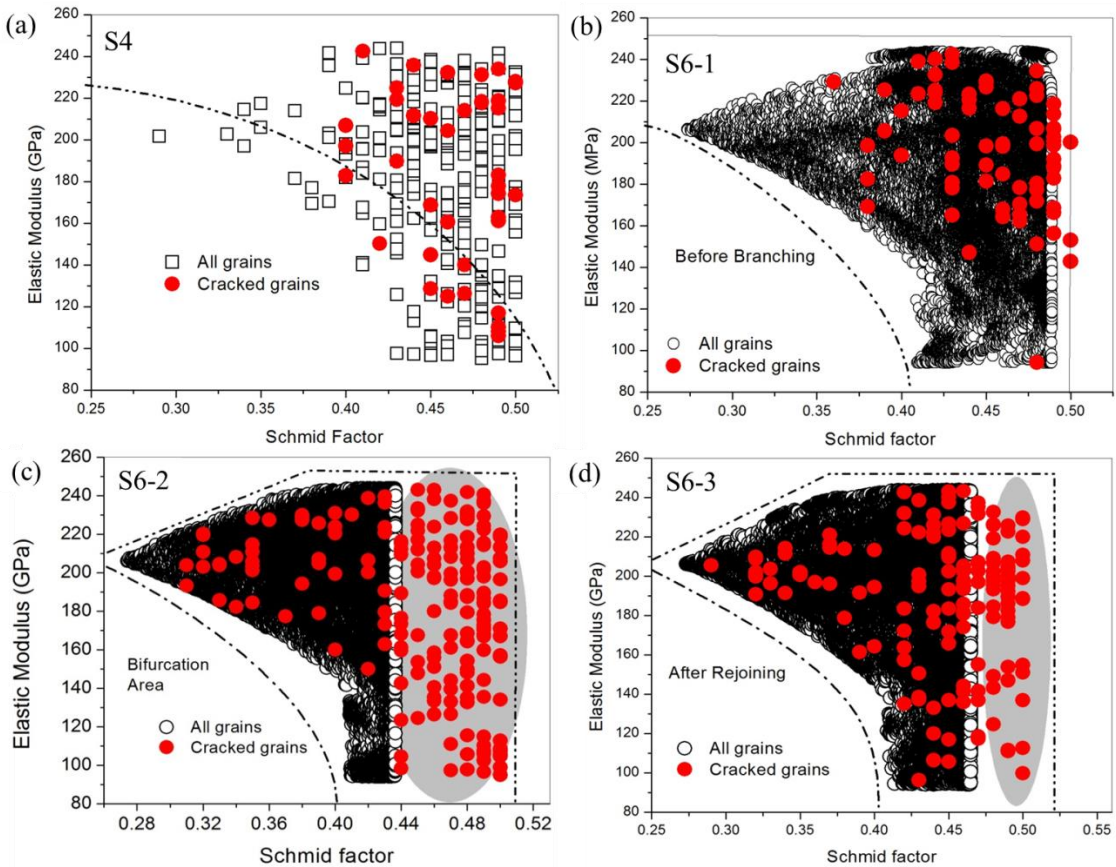

**Figure 12**: Elastic modulus vs. Schmid Factor maps of the investigated samples i.e. (a) S-4, (b) S6-1 (c) S6-2 and (d) S6-3. Distinct zones have been marks depending upon the location of the values, considering all grains and cracked grains.

In order to determine the elastic plastic incompatibility, here we have determined the elastic modulus and Schmid factors of all the grains around the crack path shown in Figure 12. It can be seen that all the cracked grains lie in the lower bound in Elastic Modulus-Schmid factor maps. The minimum Elastic modulus is shown as 200 GPa and Schmid factor is 0.4. However, some of the cracked grains are having higher Schmid factor values (0.44 to 0.52) compared to their uncracked counterparts. These high Schmid factor values denote the 'weaker' grains [42]. Larger the Schmid factor, smaller is the stress required to yield the grain in a polycrystalline ensemble. Like in any initiation part of the crack (Figure 12 (a, b)), prevalence of lower Schmid factor grains can be seen. Lower SF has been seen to increase crack path tortuosity [43]. This is because of the initial resistance offered by microstructure and smaller reversed plastic zone ahead of the crack. But, as the crack length increases i.e. as the crack advances, the plastic zone grows and the resistance of the grains got dropped [44]. Therefore, here we can see a lot of grains having distinctively higher Schmid factor than the grains existing in the uncracked portion of microstructure. The elastic modulus also having a particular range of 80 GPa to 200 GPa up to a Schmid factor of 0.4 where neither cracked nor uncracked grains have been observed. However, lower Schmid factor grains having higher elastic modulus are seen to be present in the crack path. This phenomenon can be attributed to the higher elastic mismatch with neighboring grains and favorable elastic incompatibility facilitating the cracking process.

**4.6.4 Role of grain twin pairs in crack propagation:**

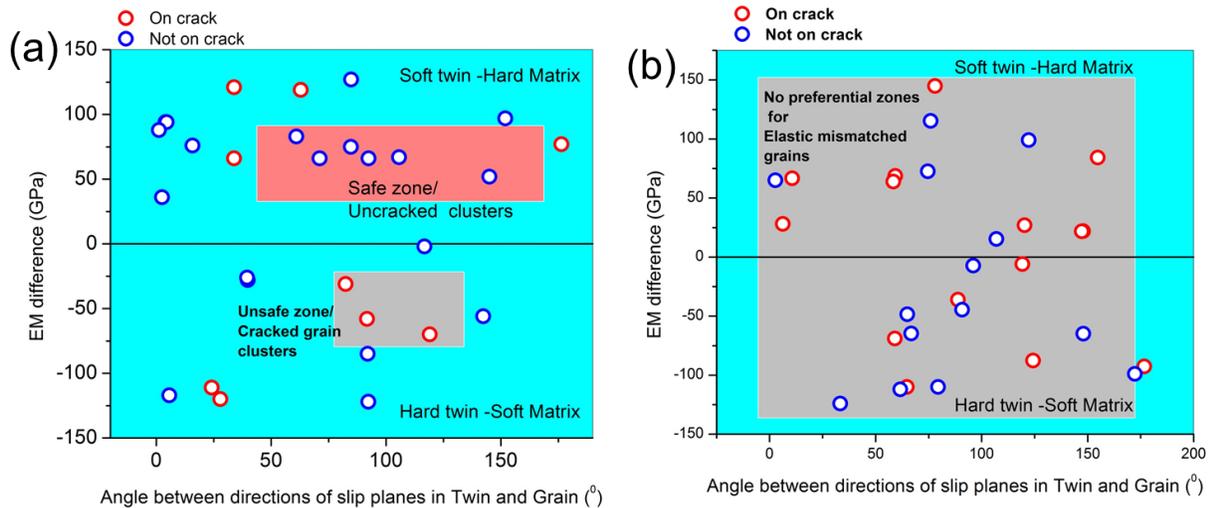

**Figure 13:** Evolution of elastic modulus difference between grain and twin data with grain twin angle (a) sample S4 and (b) sample S6.

As it has been discussed prior to this section that the dislocation mediated slip in grain twin pairs largely depends upon the interplanar angle between grain and twin. The ease of slip transfer always depends upon the preferential orientation of twin embedded in the parent grain. But these twin boundaries take part in slip blocking most of the times and thus raising the local TRSS (twin resolved shear stress) [45]. As the twin resolved shear stress is higher for the grain twin pair, instead of easy slip transfer, cracking happens. For a fatigue crack, propagating in grain twin pair, emits dislocation which are stopped or bypassed effectively in the twinned region based on the mutual orientation of matrix plane with maximum Schmid factor and twin plane. In Figure 13a, the elastic modulus difference between matrix and twin has been mapped for Sample S4 with aforementioned grain twin angle. Sometimes, the twinned region is elastically 'harder' than the matrix showing negative elastic modulus difference. There are clusters of cracked and uncracked grain-twin pairs scattered in Figure 13a. These clusters mean that for a particular combination of Elastic modulus difference and grain twin pair angle, cracking gets perturbed when especially matrix is elastically harder than the twin. The cracking process is facilitated when the twin is harder than the matrix. Therefore, the effective orientation of the twin boundary plane regarding the incoming and outgoing slip plane is having a role in the whole process. This may be governed by SF difference [46] or by Elastic modulus difference combined with other crystallographic factors [47]. An emphasis on this type of analysis has been presented in section 4.6.5. Whereas, in S6 (Figure 13b), no such decipherable trend exists. The absence of such a trend can be attributed to the fact that S6 cannot be considered as a microstructurally or physically short crack. It is indeed a long crack traversing through an aggregate of over 100 grains and governed much more by mesoscale fracture mechanics rather by local crystallography [48]. However, there can be a crystallographic initiation event involved with the cracking process in S6 but the role of grain-twin pair is absent in the progression of the crack.

**4.6.5 The slip transfer for the grain twin pairs of crack path: Luster Morris parameter**

As discussed in the previous crack initiation portion of the paper (section 4.6.1), the slip transfer problem can be addressed similarly for the subcritical crack path. As the crack propagates, the neighboring grains with favorable Schmid Factor allow the crack to grow within them. Similarly, we can tell if hard slip conditions are provoked by the local grain arrangement, the microstructural section will be resistant to crack propagation [49]. In order to understand the micromechanics behind the slip transfer, notable works from Luster and Morris [20] have been considered in the current study. Twinned grains are found to be creating hard slip conditions, whereas soft matrix grain favors slip or crack propagation.

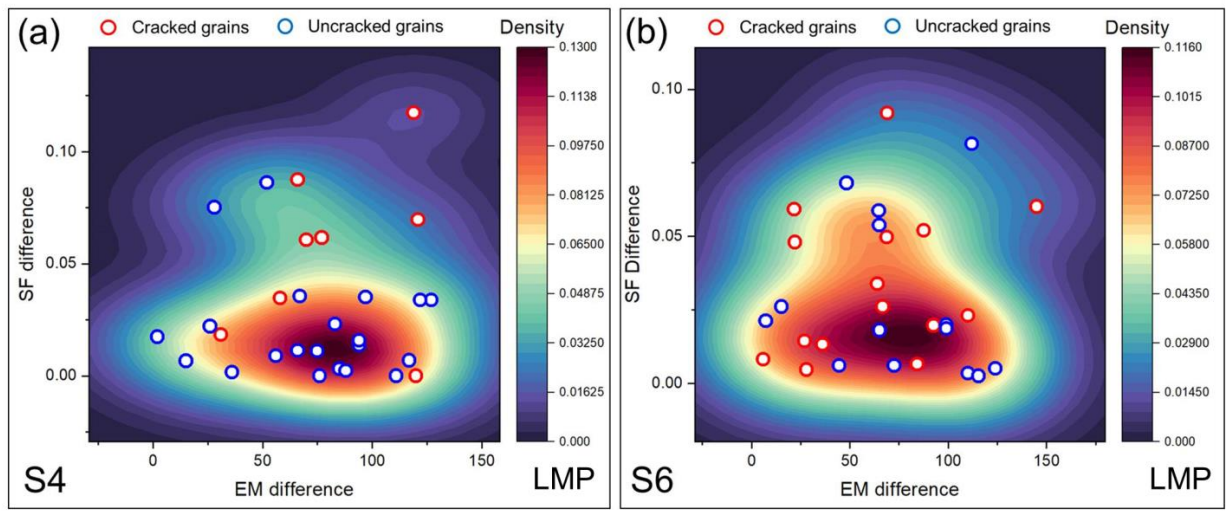

**Figure 14:** Luster Morris parameter ((LMP) as a function of Schmid factor and Elastic Modulus difference represented as contour plot, red circle represents the cracked grains and blue circles represent uncracked grain.

The likelihood of slip transfer and crack propagation always found a limited space in literature. The driving force for slip transfer and crack ($K_{cracktip}$) are interrelated. The slip lines emanating from the tip of the crack get transferred in nearby available grain and the role of twin is also very much dominant there. In our prior observation in sample S4, the Schmid factor map has been constructed around the crack in which several grain twin pairs have been seen to be involved. The twins (orange colour) present in Figure 10f are having an SF value between 0.35 to 0.38, i.e. on the medium to higher side of Schmid factor range of 0.23 to 0.5. The green grains in Figure 10f are of considerably low Schmid factor and situated far from the cracked regions. The whole phenomena of crack initiation and propagation are governed by slip transfer and the requirement of slip transfer is being fulfilled by distribution of Schmid factor. Crack propagation is a much-matured phase of strain transfer or slip transfer. Wang et al. in their phenomenal work of [50] slip transfer in commercially pure Titanium, described different types of grain pairs that mediate slip stimulated twinning. The slip stimulated twinning always refers to the twinning mode during deformation. But before deformation, if annealing twins are preexisting, how the crack interacts with those twins are one of the prime aspects of this manuscript.

The crystallographic aspects behind Luster Morris parameter have been illustrated in section 2.2.3. The orientation matrix values of neighboring grain twin pairs obtained from EBSD data are fed into an in-house spreadsheet calculator to calculate the LMP values. The grain twin pairs involved in calculations of Figure 13 has been considered. It has been assumed that the grain boundary plane is parallel to the z-axis of the calculation. So, the three aspects, SF difference at Y-axis, EM difference at X axis and LM parameter at Z-axis is represented in XY-plane contours in Figure 14.

Following are some of the pertinent observations from contour plots in Figure 14:
- In S4 sample, densities of the uncracked grains are more in the peak region of LMP with a value ~0.13 (Figure 14a). Slip transfer has been easy in grains with higher LMP and therefore the mutual ease of slip transfer does exist in uncracked grains.
- The population band of cracked grains start from the half of peak LMP value (0.065) and almost all the cracked grains lie in slip restricted regions i.e., 0.01 to 0.065 for S4.
- It has been observed in S6, that the relative density of cracked grains is higher in high LMP (0.06 to 0.12) region compared to S4 (Figure 14b). The upper bound of distribution for cracked grains is at LMP 0.12 in S6 unlike a lower value of 0.065 in S4.
- This is critical evidence of slip homogenization when the strain amplitude is raised from 0.6% (S4) to 0.8% (S6). With the increase of strain amplitude, the slip transfer demarcation becomes more diffused.

The general notion is the grains with lower Schmid factor (SF) are hard grains or the grains resistant to slip, but the grains with higher SF are the grains susceptible to slip. The Schmid factor (SF) range for the existing grains lies between 0.25 to 0.5 considering both the samples (Figure 12). So in a comparative scale, grains having SF close to 0.25 can be considered hard. Similarly, the cluster of neighboring grains with a higher difference in SF do represent the pair of soft/hard grain whereas the grain pairs with very low SF difference represent slip transfer equivalence i.e. soft/soft or hard/hard; as envisaged in Y-axis of contour plots of Figure 14. When it is pretty difficult to comment from the SF difference about the probability of slip transfer and probability of crack propagation, possibly LMP gives us a better insight. It is evident that the core of contour plot is populated by uncracked grains and the realm of the same is populated by cracked ones in S4 (Figure 14a) whereas no such demarcation does exist for S6. This may be a characteristic of oxide induced short crack in S4, while in S6, the long crack has many different features like meandering, bifurcation and rejoining. Therefore, in S6, the cracked grains are not agglomerated in a particular region rather they are sparsely distributed throughout the contour plot, Figure 14b.

### 4.6.6 Microstructure at Crack arrest location:

Figure 15a is showing the EBSD indexed zone in the crack closure region. The extracted micrographs from the EBSD analysis show an area with 73% of high angle grain boundaries (HAGB>$15^0$) among that 31% is Σ3 twin boundary. Kernel average misorientation (KAM) map of the scanned portions representing more strain in the grains on the vicinity of the main crack path, as showed by the colour legends in Figure 15b, where preferably more strain is showed by the yellow/green colour compared to the blue regions. Some strain also partitioned alongside grain boundaries. Also, it is quite evident that strain is selectively partitioned in the grain boundaries of the coarse grains (average grain size ~90 µm), Figure 15b. It is quite apparent that more strain is partitioned towards the domains nearer to the main fracture plane irrespective of the grain size; Figure 15b. Grain orientation spread (GOS) map of the crack path shown the recrystallized grain fractions far from the crack region, Figure 15c. The GOS value of less than 1° is considered to be the recrystallized fractions of any polycrystalline material. Here, it is around 75% for the current investigated region (showed by bluish colour) which is in line with the observations from KAM map.

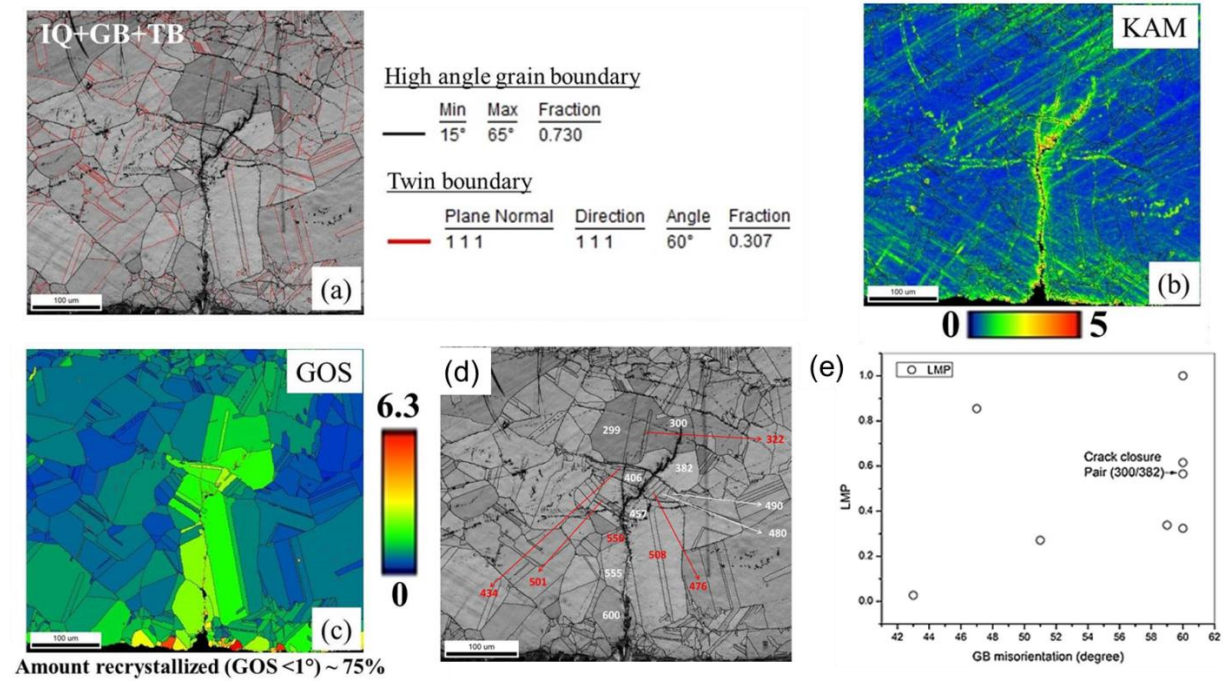

**Figure 15**: EBSD scanned images of the crack closure grains. (a) Image quality map embedded with low angle and high angle grain boundary map (IQ+GB), (b) kernel average misorientation (KAM) and (c) grain orientation spread (GOS) map of the same. (d) Grain-grain and grain twin pairs involving crack closure (e) Luster Morris parameter involving crack arrest.

**Table 3:** Luster-Morris Parameter (LMP) on crack arrest grains

| Grain / Twin 1 | Phi 1 | Phi | Phi 2 | Grain / Twin 2 | Phi 1 | Phi | Phi 2 | GB (Degree) | LMP |
|---|---|---|---|---|---|---|---|---|---|
| 600 | 246.8 | 50.8 | 37.8 | 555 | 142.1 | 29.3 | 80.3 | 51 | 0.271023 |
| 555 | 142.1 | 29.3 | 80.3 | 550 | 259.8 | 29.4 | 6.2 | 60 | 0.615568 |
| 457 | 43.6 | 39.2 | 26.1 | 406 | 236.9 | 38.9 | 36.5 | 47 | 0.854539 |
| 508 | 137.5 | 28.9 | 84.4 | 457 | 43.6 | 39.2 | 26.1 | 60 | 0.323162 |
| 480 | 216.2 | 45.1 | 48 | 476 | 41.8 | 26 | 40.7 | 60 | 0.999876 |
| 382 | 345.8 | 32.8 | 72.7 | 300 | 206.9 | 45.4 | 23.2 | 60 | 0.56569 |
| **406** | **239** | **39.2** | **36** | **299** | **291.1** | **28.8** | **3.9** | **43** | **0.027087** |
| 490 | 128.1 | 15.7 | 75.6 | 480 | 216.2 | 45.1 | 48 | 59 | 0.33797 |

From Figure 15(d, e), an attempt has been made to correlate the Luster Morris parameter (LMP) with GB misorientation. Figure 15d represents the IQ map with grain numbering and Figure 15e represents the GB misorientation with LMP. The idea behind this correlation is to observe how the mutual orientation of two grains does affect the slip transferability. Grain boundary misorientation represents the local geometry along misorientation axis and definitely influence the transmission of slip. Figure 15e does represent that the grain twin pair with GB misorientation 60° is responsible for crack arrest whereas a LMP of 0.6 denotes slip transferability leaning on the easy slip side. But prior to that the slip transferability between grain 406 and 299 is very low (~0.027) as envisaged from Table 3. This non-coplanar configuration is a reason leading to crack arrest.

Generally in FCC system, the crack arrest takes place in the high indexed planes of the arresting grain [51]. But, if we see from the concept of slip transfer, this can result in a different qualifying criterion for arrester grains. Fracture mechanics involve dislocation emission from the crack tip and it is an inevitable process. The slip lines emanating from the crack tip are expected to get transferred from one grain to another depending upon the microstructural elasto-plastic field. This elasto-plastic field is deeply entangled with microstructural features like GB/TB characteristics, retained stress in grain interior, mutual orientation of the individual grain and ease of slip transfer in the grain aggregate. The intensity of this field is expected to depend upon the ease of slip transfer, quantitatively denoted by LMP. Here, the LMP between grain 406 and 299 and 299 and 300 are exceptionally low and these low LMP values activate the slip hardening as well as increased fracture resistance increase thereof.

## 5. Conclusive remarks:

In the present study, secondary cracks of superalloy Haynes 282 generated after HTLCF have been characterized for stability, mode mixicity and crystallographic aspects. This is the first ever report on an integrated approach to corelate slip activity, elasto-plastic incompatibility and grain boundary geometry. From this study, the following conclusions can be drawn

- At low strain amplitudes (0.4%), constant CTOA portion in stable crack regime is more because of longer crack propagation time for all temperature domain. At higher temperatures (i.e. 700 and 760ºC) and higher strain amplitudes (0.6% and 0.8%), the length of non-constant regime is greater in some cases because of higher CTOA due to decreased resistance to fracture.
- Mode mixicity has been found to decrease with strain amplitudes for furnace cooled microstructures but it is showing an opposite trend for air cooled samples. Based on our observation we can apprise that the change in mode mixicity can be expressed as a function of microstructures.
- Different stages of crack initiation and propagation have been identified and a mechanism map has been constructed under high temperature low cycle fatigue conditions. Detailed meso mechanisms have been superimposed on CTOA vs. crack length plot.
- A grain cluster having a preferential orientation for slip transfer from one grain to another has been located under the facet plane. This grain cluster acts as a crack initiator for the investigated crack path.
- From the estimation of the elasto-plastic mismatch, it has been established that higher elastic mismatch is the dominant cause facilitating LCF crack propagation.
- The positive elastic mismatch between twin and grain pairs for short crack exhibit a particular crack resistant cluster depending upon their mutual angular orientation beyond 55°. Vis-a vis, a negative elastic mismatch between twin and grain, gives rise to a cracked cluster. However, this observation does not hold true for long cracks. Therefore, it can be established that the mutual angular orientation between grain and twin plane is also a critical microstructural constraint for crack propagation, especially for short cracks.
- Detailed slip transfer studies depict slip homogenization with increasing strain amplitude and vividly low LMP value less than 0.03 results into crack arrest. High LMP denotes mutual ease of slip transfer which do exist in uncracked grain twin pairs in S4. Grain-Twin pairs with higher elastic and plastic mismatch facilitate crack propagation in short crack regime i.e. S4. Whereas, in long crack regime, i.e. S6, mesoscale fracture parameters ignore the crystallographic effects.
- In crack arrest microstructure, the grain cluster containing twin boundary resists crack propagation where the mutual ease of slip transfer between two grains does not exist because of remarkably low LMP value (<0.03). In the investigated crack arrest microstructure, non-coplanar mutual existence of grains is the effective crack arrest configuration.

**Acknowledgement:**

We are very much grateful to staffs, NML creep bay laboratory for their help in producing the data. Also we thank Dr. S. Srikanth, Director, NML and Dr. S. Tarafder for their technical input and throughout support from the inhouse grant OLP 0161.